\documentclass[amsmath,amssymb,aps,letterpaper,prd,twocolumn,longbibliography,10pt]{revtex4-1}
\pdfoutput=1
\usepackage{graphicx}
\usepackage{color}
\usepackage{comment}

\usepackage{subfigure}

\usepackage{color}

\makeatletter \let\@keywords\@empty \let\@subject\@empty
\providecommand{\keywords}[1]{\gdef\@keywords{#1}}
\providecommand{\subject}[1]{\gdef\@subject{#1}}
\def\thetitle{\@title}
\def\theauthor{\@author}
\def\thesubject{\@subject}
\def\thedate{\@date}
\def\thekeywords{\@keywords}
\makeatother
\AtBeginDocument{%
\hypersetup{pdftitle={\thetitle}}%
\hypersetup{pdfauthor={\theauthor}}%
\hypersetup{pdfsubject={\thesubject}}%
\hypersetup{pdfkeywords={\thekeywords}}%
}

\usepackage[bookmarks=true,hyperfigures=true]{hyperref}
\usepackage{fixmath}
\usepackage{mathtools}

\providecommand{\href}[2]{#2}

\let\oldbfseries=\bfseries
\let\oldmdseries=\mdseries
\let\oldnormalfont=\normalfont
\renewcommand{\bfseries}{\oldbfseries\boldmath}
\renewcommand{\mdseries}{\oldmdseries\unboldmath}
\renewcommand{\normalfont}{\oldnormalfont\unboldmath}

\makeatletter
\newlength{\apb@width}
\newcommand{\autoparbox}[2][c]{\settowidth{\apb@width}{#2}\parbox[#1]{\apb@width}{#2}}

\makeatother

\newcommand{\beq}{\begin{equation}}
\newcommand{\eeq}{\end{equation}}
\newcommand{\bea}{\begin{eqnarray}}
\newcommand{\eea}{\end{eqnarray}}

\newcommand{\CL}{\mathcal{L}}
\newcommand{\CO}{\mathcal{O}}

\newcommand{\spa}{\ , \ \ }
\newcommand{\ds}{\displaystyle}

\begin{document}

\title{\Large On the existence of the Blandford-Znajek monopole \\[1mm] for a slowly rotating Kerr black hole}

\author{Gianluca Grignani$^1$}%
 \email{grignani@pg.infn.it}
\author{Troels Harmark$^2$}%
 \email{harmark@nbi.ku.dk}
\author{Marta Orselli$^{1,2}$}%
  \email{marta.orselli@pg.infn.it}

\affiliation{%
$^1$Dipartimento di Fisica e Geologia, Universit\`a di Perugia, I.N.F.N. Sezione di Perugia, \\ Via Pascoli, I-06123 Perugia, Italy \\
$^2$Niels Bohr Institute, Copenhagen University,  Blegdamsvej 17, DK-2100 Copenhagen \O{}, Denmark
}%

\begin{abstract}
The Blandford-Znajek monopole is a conjectured solution of force-free electrodynamics in the vicinity of a slowly rotating Kerr black hole, supposedly defined as a perturbation in small angular momentum. 
It is used to argue for the extraction of energy from rotating black holes by the Blandford-Znajek process. We set up a careful analysis of the perturbative definition of the Blandford-Znajek monopole, showing in particular that the regime in which it is defined allows to use the technique of matched asymptotic expansions. Our conclusion is that the Blandford-Znajek monopole, as it is defined, is not consistent with demanding physically reasonable boundary conditions far away from the event horizon. This puts into question the existence of the Blandford-Znajek monopole, at least in the limit of slow rotation of the Kerr black hole.
\end{abstract}

\maketitle

\section{Introduction}

In their seminal paper \cite{Blandford:1977ds}, Blandford and Znajek proposed a mechanism that can drive the jets and gamma-ray burst observed from Active Galactic Nuclei and stellar black holes. They showed that in the vicinity of a black hole the electromagnetic field is force-free, in the sense that it decouples from other degrees of freedom. The equations for force-free electrodynamics (FFE) are
\begin{equation}
\label{FFE_eqs}
\begin{array}{c}\ds
F_{\mu \nu}=\partial_{\mu} A_{\nu}-\partial_{\nu}A_{\mu} \spa
D_\mu  F^{\mu\nu} = - J^\nu \,, \\[2mm] \ds  F_{\mu\nu} J^\nu = 0 \spa J^\mu \neq 0 \,,
\end{array}
\end{equation}
where $A_\mu$ is the gauge potential, $F_{\mu\nu}$ the electromagnetic field strength and $J^\mu$ the current.
These are basically the ordinary Maxwell's equations with the additional requirements that the electromagnetic energy-momentum tensor is conserved and that the current is non-zero. 
The accretion disc surrounding the black hole acts as a magnetic source. For a rotating black hole the frame-dragging of the event horizon provides a type of magnetic Penrose process - known as the Blandford-Znajek process -  that allows the  extraction of rotational energy from the black hole \cite{Blandford:1977ds,Komissarov:2008yh,Lasota:2013kia}.

Blandford and Znajek developed an analytic description of this process by finding a perturbative solution of the FFE equations \eqref{FFE_eqs} around a Kerr black hole. The simplest of these solutions is the so-called Blandford-Znajek monopole that easily can be turned into a split-monopole solution with a magnetic source at a disc modelling an accretion disc around the black hole. The Blandford-Znajek monopole is defined as a perturbation in the rotation parameter $\alpha = J/(GM^2)$ of the Kerr black hole, with $J$ being the angular momentum and $M$ the mass. For $\alpha=0$ it is a static magnetic monopole. At order $\alpha$ the monopole is rotating with the event horizon, and the solution resembles the Michel monopole \cite{1973ApJ...180L.133M} with the angular velocity being half the one of the Kerr black hole. At order $\alpha^2$ one finds then a non-trivial correction to the magnetic monopole configuration.

An issue that already was pointed out as a difficulty in the original paper \cite{Blandford:1977ds} is what boundary conditions one should demand in the asymptotic region, {\sl i.e.}~far away from the black hole. Writing $r$ as the radial coordinate in the black hole space-time with $r=r_+$ being the location of the event horizon, the asymptotic region is $r \gg r_+$. Only recently this issue has been studied in more detail \cite{Tanabe:2008wm,Pan:2015haa}. In particular, one of the goals of this work is to understand better the apparent divergency for $r\rightarrow \infty$ in the perturbative solution for $F_{\mu\nu}$ at order $\alpha^4$ as compared to the zeroth order solution (order $\alpha^0$) \cite{Tanabe:2008wm}.

We argue that the correct way to address the issue of boundary conditions in the asymptotic region $r \gg r_+$ is in terms of the technique of matched asymptotic expansions \cite{Poisson:2003nc}. This technique gives a framework for finding perturbative solutions in a black hole space-time where one has to satisfy boundary conditions both at the event horizon as well as in the asymptotic region. It illuminates that the expansion of the FFE equations \eqref{FFE_eqs} for small $\alpha$ only works sufficiently near the event horizon. In the asymptotic region, one has to  expand the FFE equations in $1/r$  instead
\footnote{This expansion was recently considered in \cite{Li:2017qzu} although without using the technique of matched asymptotic expansions to connect it to the near-horizon region.}.  

Since the Blandford-Znajek monopole is defined as a perturbation in $\alpha$, we demand that the solution can be made arbitrarily small at order $\alpha^n$ compared to the previous order $\alpha^{n-1}$ by choosing a sufficiently small $\alpha$. From this we can infer our boundary condition for $F_{\mu\nu}$ in the asymptotic region.

Using the framework of matched asymptotic expansions we show that imposing regularity of $F_{\mu\nu}$ at the event horizon along with our boundary condition in the asymptotic region leads to inconsistencies already at order $\alpha^2$, {\sl i.e.}~the first order in which one has a correction to the Michel monopole. Since we have imposed only the boundary conditions in the asymptotic region that follows from demanding that the perturbation in $\alpha$ does not blow up in the asymptotic region, we are forced to conclude that the Blandford-Znajek monopole, as defined perturbatively in $\alpha$, does not exists.

Note that our analysis could potentially have resolved the divergency found at order $\alpha^4$ in  \cite{Tanabe:2008wm} since their result in any case breaks down for $r \rightarrow \infty$ since, as already remarked, one cannot use the FFE equations expanded in $\alpha$ for $r\rightarrow \infty$. Instead our analysis points to a problem in the asymptotic region already at order $\alpha^2$.

There are several papers studying the Blandford-Znajek monopole numerically 
\cite{Komissarov:2001sjq,Komissarov:2004ms,McKinney:2004ka,McKinney:2006sc,Tchekhovskoy:2009ba,Palenzuela:2010xn,Contopoulos:2012py,Nathanail:2014aua,Mahlmann:2018ukr}.  These papers claim to find the Blandford-Znajek monopole for finite values of $\alpha$. It would be highly interesting to understand better how to reconcile these numerical studies with the conclusions of this paper. We comment further on this in the conclusions. 

\section{Force free electrodynamics around Kerr black hole}

We begin by reviewing the equations for FFE in the background of a Kerr black hole following the expositions in \cite{McKinney:2004ka,Tanabe:2008wm,Pan:2015haa}.
The metric for the Kerr black hole in Kerr-Schild coordinates is
\begin{eqnarray}
&&ds^2=-\left(1-\frac{r_0 r}{\Sigma}\right)dt^2+\left(\frac{2r_0 r}{\Sigma}\right)dr dt+\left(1+\frac{r_0 r}{\Sigma}\right)dr^2 \cr
&&+\Sigma d\theta^2-\frac{2 a r_0 r \sin^2{\theta}}{\Sigma}d\phi dt 
-2a\left(1+\frac{r_0 r}{\Sigma}\right)\sin^2\theta d\phi dr \cr
&&+\left(\Delta+\frac{r_0 r(r^2+a^2)}{\Sigma}\right)\sin^2\theta d\phi^2 \,,
\label{metric}
\end{eqnarray}
where we defined
\begin{eqnarray}
&&\Sigma = r^2+a^2 \cos^2 \theta \spa  \Delta=(r-r_+)(r-r_-)\cr
&&  r_\pm = \frac{1}{2} r_0 \pm \sqrt{\frac{r_0^2}{4} - a^2}  \spa 
r_0=2G M \spa a=\frac{J}{M}.
\end{eqnarray}
Here $G$ is Newtons constant while $J$ and $M$ are the angular momentum and mass of the Kerr black hole, respectively.
The Kerr black hole is stationary and axisymmetric corresponding to the commuting Killing vector fields $\partial_t$ and $\partial_\phi$. We consider here FFE configurations that are stationary and axisymmetric around the same rotation axis as the Kerr black hole. The FFE equations are \eqref{FFE_eqs}. By demanding stationarity and axisymmetry we can choose a gauge in which $\partial_t A_\mu = \partial_\phi A_\mu = 0$. 
We define the magnetic flux function
\begin{equation}
\psi(r,\theta)=A_\phi(r,\theta) \,.
\end{equation}
This is the magnetic flux flowing upwards through a circular loop centered around the rotation axis and going through the point $(r,\theta)$. Note that we impose the condition $\psi=0$ at $\theta=0$ as there are no sources at the rotation axis.

From \eqref{FFE_eqs} we see that $F_{\mu\nu} D_\rho F^{\nu\rho} = 0$. Combining the $\mu=t,\phi$ components of this equation we get $\partial_r A_t \partial_\theta \psi = \partial_\theta A_t \partial_r \psi$ which means one can regard $A_t$ as a function of $\psi$. Using this we define $\Omega(r,\theta)$ by
\begin{equation}
\label{def_omega}
\partial_r A_t = - \Omega \, \partial_r \psi \spa \partial_\theta A_t = - \Omega \, \partial_ \theta \psi \,.
\end{equation}
$\Omega$ is the angular velocity of the magnetic field line.
From Eq.~\eqref{def_omega} one can infer 
\begin{equation}
\label{intcond1}
\partial_r \Omega \, \partial_\theta \psi = \partial_\theta \Omega\, \partial_r \psi \,,
\end{equation}
which means that $\Omega$ can be regarded as a function of $\psi$. Thus, Eq.~\eqref{intcond1} is an integrability condition for $\Omega$. 
Define furthermore
\begin{equation}
I = \sqrt{-g} F^{\theta r} \spa B^\phi = \frac{1}{\sqrt{-g}} F_{r\theta} \,.
\end{equation}
Here $I$ is the total electric current flowing upward through the $(r,\theta)$ loop and $B^{\phi}$ is the toroidal magnetic field.
From the $\mu=t,\phi$ components of $F_{\mu\nu} D_\rho F^{\nu\rho} = 0$ one also finds
\begin{equation}
\label{intcond2}
\partial_r I \, \partial_\theta \psi = \partial_r \psi \, \partial_\theta I \,,
\end{equation}
which is the integrability condition for $I$ ensuring that it can be regarded as a function of $\psi$. 
From the $\mu=r,\theta$ components of $F_{\mu\nu} D_\rho F^{\nu\rho} = 0$ one finds instead the {\sl Stream equation}
\begin{equation}
\label{stream}
-\Omega\, \partial_\mu (\sqrt{-g}F^{t \mu})+\partial_\mu (\sqrt{-g}F^{\phi \mu})+F_{r \theta}\frac{dI}{d\psi} = 0 \,.
\end{equation}
This is a non-linear equation that relates the three functions $\psi$, $\Omega$ and $I$. Finally, we find that the toroidal magnetic field is given by 
\begin{equation}
\label{genBphi}
B^{\phi}=-\frac{I \Sigma+\big(\Omega r_0 r-a\big) \sin \theta \partial_{\theta}\psi}{\Delta \Sigma \sin^2\theta}.
\end{equation}

Finding a solution of the FFE equations corresponds to finding $\psi$, $\Omega$ and $I$ that solves the integrability conditions \eqref{intcond1} and \eqref{intcond2} as well as the Stream equation \eqref{stream}. At the event horizon   $r=r_+$ of the Kerr black hole we demand regularity of $\psi$ and $B^\phi$. That $B^\phi$ is regular at the horizon is equivalent to the Znajek condition in Boyer-Lindquist coordinates \cite{1977MNRAS.179..457Z}. The conditions in the asymptotic region $r\rightarrow\infty$ are discussed later in the paper.

\section{Blandford-Znajek monopole from perturbative expansion in $\alpha$}

The Blandford-Znajek (split-)monopole is defined perturbatively as follows \cite{Blandford:1977ds, McKinney:2004ka,Tanabe:2008wm,Pan:2015haa}. One starts with a static (split-)monopole solution of the FFE equations \eqref{FFE_eqs} with $\Omega=I=0$ in the background of the Schwarzschild black hole. Thus, neither the background geometry, nor the FFE fields are rotating. Then one considers turning on a small rotation both for the black hole background, as well as for the FFE fields at the same time. One makes this expansion in the dimensionless parameter $\alpha=J/(GM^2)=2a/r_0$ proportional to the angular momentum of the Kerr black hole.

We note that this method to find solutions to the FFE equations for a slowly rotating Kerr black hole has been employed in several other cases, all starting with a static FFE solution in the background of the Schwarzschild black hole. This includes the parabolic solution \cite{Blandford:1977ds}, the vertical uniform solution \cite{Pan:2014bja} and the hyperbolic solution \cite{Gralla:2015vta}. 

One can argue that since $\psi(r,\theta)$ is related to the shape of the magnetic field lines, the perturbative expansion of $\psi(r,\theta)$ is naturally in even powers of $\alpha$ so that changing the sign of the angular momentum does not alter the shape of the field lines \cite{Blandford:1977ds}. Since one wants the expansion of $\Omega(\psi)$ and $I(\psi)$ to start at order $\alpha$, one gets that these expansions  are in odd powers of $\alpha$. This in turn means that $B^\phi$ is expanded in odd powers of $\alpha$ as well.

Hence, we expand the four fields $\psi$, $\Omega$, $I$ and $B^\phi$ in powers of $\alpha$ as follows
\begin{equation}
\label{alpha_exp}
\begin{array}{c} \ds
\psi (r,\theta)= \psi_{(0,\cdot)} + \alpha^2 \psi_{(2,\cdot)} + \alpha^4 \psi_{(4,\cdot)} + \CO(\alpha^6) \,,
\\[2mm]\ds
r_0 \Omega (r,\theta) = \alpha\Omega_{(1,\cdot)}+\alpha^{3} \Omega_{(3,\cdot)} + \CO(\alpha^5)\,,
\\[2mm]\ds
r_0 I  (r,\theta)=  \alpha I_{(1,\cdot)} + \alpha^{3} I_{(3,\cdot)}+ \CO(\alpha^5)\,,
\\[2mm]\ds
B^\phi (r,\theta)=  \alpha B_{(1,\cdot)} + \alpha^{3} B_{(3,\cdot)}+ \CO(\alpha^5)\,,
\end{array}
\end{equation}
where $\psi_{(m,\cdot)}$, $\Omega_{(m,\cdot)}$ and $I_{(m,\cdot)}$ are functions of $r$ and $\theta$. One should impose \eqref{intcond1} and \eqref{intcond2} order-by-order in $\alpha$ to ensure that $\Omega=\Omega(\psi)$ and $I=I(\psi)$. The reason for our notation $(m,\cdot)$ will be clear below. 

One can now expand the Stream equation \eqref{stream} in terms of the expansions \eqref{alpha_exp}. 
More specifically, we expand the field $\psi$ in even powers of $\alpha$ and the fields $\Omega$, $I$ and $B^\phi$ in odd powers in $\alpha$ and put this into the Stream equation \eqref{stream} along with the metric \eqref{metric}, and subsequently expand the Stream equation in powers of $\alpha$.
Schematically, this gives the equations \cite{Blandford:1977ds, McKinney:2004ka,Tanabe:2008wm,Pan:2015haa}
\begin{equation}
\mathcal{L}\psi_{(2m,\cdot)}(r,\theta)=S_{2m}(r,\theta)\,,
\label{orderstream}
\end{equation}
for $m=0,1,...$,
where $\mathcal{L}$ is the differential operator
\begin{equation}
\mathcal{L}\equiv\frac{1}{\sin\theta}\partial_r\left(1-\frac{r_0}{r}\right)\partial_r+\frac{1}{r^2}\partial_{\theta}\frac{1}{\sin\theta}\partial_{\theta}\,,
\label{lop}
\end{equation}
and $S_{2m}$ are source terms with $S_0=0$. To zeroth order in $\alpha$ the stream equation \eqref{stream} is thus simply $\mathcal{L}\psi_0(r,\theta)=0$. This is the above-mentioned starting point where one is solving the FFE equations \eqref{FFE_eqs} with $\Omega=I=0$ in the background of the Schwarzschild black hole. 
The solution that we are interested in is the static monopole \cite{Blandford:1977ds}
\begin{equation}
\psi_{(0,\cdot)}=1-\cos\theta \,.
\label{zero}
\end{equation}
One can instead use the static split-monopole as the starting point. This is obtained by using the monopole of opposite charge $\psi_{(0,\cdot)}=-(1-\cos\theta)$  for $\pi/2 < \theta \leq \pi$ which is below the plane at $\theta=\pi/2$. The static split-monopole is an exact solution of Maxwell's equations except on a current sheet located at the plane $\theta=\pi/2$ \cite{Blandford:1977ds, Gralla:2014yja}. In the rest of the paper we will restrict ourselves to the northern hemisphere.
One can trivially extend all of the computations of the monopole case below to the case of the split-monopole by changing signs for $\pi/2 < \theta \leq \pi$. For simplicity, we consider only the monopole below.

To first order in $\alpha$ one gets from \eqref{intcond1} and \eqref{intcond2} that $\Omega_{(1,\cdot)}$ and $I_{(1,\cdot)}$ are functions only of $\theta$. Inserting this in \eqref{genBphi} we get
\begin{equation}
B_{(1,\cdot)}^\phi = \frac{r_0 - 2r \Omega_{(1,\cdot)}(\theta) -  \frac{2r^2I_{(1,\cdot)}(\theta)}{r_0\sin^2 \theta}}{2r^3 (r-r_0)}\,.
\end{equation}
Demanding that $B^\phi$ should be regular at the event horizon one finds
\begin{equation}
\label{alpha_one}
I_{(1,\cdot)}(\theta)=\frac{1-2\Omega_{(1,\cdot)}(\theta)}{2} \sin^2\theta\,.
\end{equation}
This gives
\begin{equation}
\label{B_first}
B_{(1,\cdot)}^\phi = - \frac{1-2\Omega_{(1,\cdot)}(\theta)}{2r_0r^2} - \frac{1}{2r^3} \,.
\end{equation}

To second order in $\alpha$ one gets a non-zero source term $S_2$ that depends on $\Omega_{(1,\cdot)}(\theta)$. The dominant terms for $r\rightarrow \infty$ are
\begin{equation}
\label{S2gen}
S_2 = \frac{4\Omega_{(1,\cdot)} - 1}{2r_0^2} \sin \theta \cos \theta + \frac{\frac{d}{d\theta}\Omega_{(1,\cdot)}}{2r_0^2} \sin^2 \theta + \CO (r^{-2})\,.
\end{equation}
If $S_2$ goes like $r^0$ for $r\rightarrow \infty$ it is straightforward to show from \eqref{orderstream} that  $\psi_{(2,\cdot)}(r,\theta)$ goes like $r^2$. A necessary condition for the perturbative solution to be valid is that it remains a small perturbation for small $\alpha$ everywhere. This means one cannot have that $\psi(r, \theta)$ diverges for $r \rightarrow \infty$. Demanding $S_2\rightarrow 0$ for $r\rightarrow \infty$ fixes uniquely 
\begin{equation}
\label{Omega_1cdot}
\Omega_{(1,\cdot)}=\frac{1}{4} \spa I_{(1,\cdot)} = \frac{1}{4} \sin^2\theta\,.
\end{equation}
One finds then
\begin{equation}
\label{fixedS2}
\mathcal{L}\psi_{(2,\cdot)}=-\frac{r_0}{2r^3}\left(1+\frac{r_0}{r}\right)\sin\theta\cos\theta\,.
\end{equation}
This is solved by \cite{Blandford:1977ds}
\begin{equation}
\label{a2}
\psi_{(2,\cdot)}=R(r)\sin^2\theta\cos\theta\,,
\end{equation}
with 
\begin{eqnarray}
\label{r}
&&R(r) =\frac{r_0^2+6r_0r-24r^2}{12r_0^2} \log \left(\frac{r}{r_0}\right)+\frac{11}{72}+\frac{r_0}{6r}+\frac{r}{r_0}-\frac{2r^2}{r_0^2}\cr
&&  +\left[\mbox{Li}_2\left(\frac{r_0}{r}\right)-\log \left(\frac{r}{r_0}\right) \log \left(1-\frac{r_0}{r}\right)\right]\frac{r^2(4r-3r_0)}{2r_0^3}\,.
\end{eqnarray}
This solution is finite at the horizon for $r= r_+$. Asymptotically for $r\rightarrow \infty$ we have
\begin{equation}
\label{A2_larger}
\psi_{(2,\cdot)}=\frac{r_0}{8r} \sin^2 \theta \cos \theta + \CO\Big(\frac{r_0^2}{r^2}\log \frac{r}{r_0}\Big)\,.
\end{equation}
With this result it was concluded in \cite{Blandford:1977ds} that the rotating Michel monopole is an approximate solution of the FFE equations for Kerr magnetosphere up to second order in the spin parameter $\alpha$. However, as we shall see below, this is not consistent. Indeed, we shall show that demanding that the perturbative solution is consistent for $r\rightarrow \infty$ leads to a contradiction at order $\alpha^2$.

One can continue to higher orders in $\alpha$, as was done in \cite{Tanabe:2008wm}\footnote{In \cite{Pan:2015haa} the same computation as in \cite{Tanabe:2008wm} was performed up to the 4th order in $\alpha$ by means of a convergence condition to be imposed at $r=\infty$ and the result for $\psi_{(4,\cdot)}$ was claimed to go to zero for $r\to\infty$ in accordance with the perturbation scheme. We were able to perform the computation following \cite{Pan:2015haa} but our result confirms the findings of Ref.~\cite{Tanabe:2008wm}.}. At order $\alpha^3$ one gets $\Omega_{(3,\cdot)}$ and $I_{(3,\cdot)}$ up to an undetermined function of $\theta$ which can be fixed at order $\alpha^4$ requiring that the source term $S_4\to 0$ for $r\rightarrow \infty$. This gives
\begin{equation}
\label{Omega_3cdot}
\Omega_{(3,\cdot)}=\frac{1}{16} +\frac{(67-6\pi^2)}{576}\sin^2\theta \spa I_{(3,\cdot)} = \Omega_{(3,\cdot)}\sin^2\theta\,,
\end{equation}
where we also assumed regularity at the rotation axis.
Using these results one finds a source term $S_4$ that, for $r\rightarrow \infty$, goes like 
\begin{equation}
\label{S4lead}
S_4 = -\frac{3}{64 r_0 r}\cos\theta\sin^3\theta + \CO (r^{-2})\,.
\end{equation}
With a source term that goes like $1/r$ for $r\rightarrow \infty$, it is easy to show that $\psi_{(4,\cdot)}(r,\theta)$ goes like $r$ and this invalidates the perturbative scheme.

\section{Can one save the Blandford-Znajek monopole?}

\subsection{Why the perturbation in $\alpha$ fails}

There are two possible reasons that one runs into inconsistencies when considering the perturbation theory in $\alpha$ for the Blandford-Znajek monopole solution:
\begin{itemize}
\item[1.] The Blandford-Znajek monopole solution does not exist for small $\alpha$. Thus, one runs into divergencies at order $\alpha^4$ since one is trying to approach a solution that does not exist. 
\item[2.] The Blandford-Znajek monopole solution exists for small $\alpha$. But one runs into divergencies at order $\alpha^4$ because one is not using the right approach to find the solution for $r\rightarrow \infty$.
\end{itemize}
In the following we adopt the assumption that it is the second option that is correct, {\sl i.e.}~that the Blandford-Znajek monopole solution exists, which means that one needs to take a closer look at what happens for $r\rightarrow \infty$. We shall see below that this assumption runs into inconsistencies that eventually will force us to conclude that the first option is the correct one. In a forthcoming paper \cite{New} we will present further evidence for this. 

Assuming the Blandford-Znajek monopole solution exists, it should obey the following conditions
\begin{equation}
\label{conditions}
\begin{array}{c}\ds
\psi,~ \Omega,~ I ~{\text{are finite for}} ~\frac{r}{r_0}\rightarrow \infty \,, \\[3mm] \ds
\psi ~{\rm and} ~B^\phi ~{\text{are regular at}}~ r=r_+ \,, \\[3mm] \ds
\psi~{\rm and} ~B^\phi ~{\text{are regular at lightsheets}}\,.
\end{array}
\end{equation}
For $\psi(r,\theta)$ the reason for the condition of finiteness for $r/r_0 \rightarrow \infty$ is that $\psi_{(0,\cdot)}(r,\theta)$ is finite for $r/r_0 \rightarrow \infty$. Hence the perturbations in $\alpha$ should be finite as well, as already argued above. For $\Omega$ and $I$ one finds that they are finite for $r/r_0 \rightarrow \infty$ at order $\alpha$, and hence the same reasoning applies. Another argument for the condition of finiteness for $r/r_0 \rightarrow \infty$ is that one imagines that the small $\alpha$ Blandford-Znajek monopole has some resemblance to the Michel monopole \cite{1973ApJ...180L.133M}.

Obviously, the divergence of $\psi_{(4,\cdot)}(r,\theta)$ is inconsistent with the conditions \eqref{conditions}. However, assuming the Blandford-Znajek monopole solution exists for small $\alpha$, a reason for this could be that there is an order-of-limits problem between taking $\alpha\rightarrow 0$ and $r/r_0 \rightarrow \infty$. We can see this already in Eqs.~\eqref{orderstream}-\eqref{lop}. Acting with the operator $\CL$ of Eq.~\eqref{lop} on $\alpha^m \psi_{(m,\cdot)}$ should produce a term at order $\alpha^m$. However, if $r_0/r \ll \alpha^k$ a part of $\CL (\alpha^m \psi_{(m,\cdot)})$ would become of order $\alpha^{m+k}$ and one would thus have a mixing between terms of different orders in $\alpha$. Thus, the perturbations in $\alpha$ set up by  Eqs.~\eqref{orderstream}-\eqref{lop} are only consistent in the region
\begin{equation}
\label{NH_region}
r_+ \leq r \ll \frac{r_0}{\alpha}
\end{equation}
Nevertheless, one still needs to connect the solution to the asymptotic region, to ensure that the conditions \eqref{conditions} are obeyed. If the Blandford-Znajek monopole solution exists for small $\alpha$, this should be possible to do using the method of matched asymptotic expansions \cite{Poisson:2003nc}.

\subsection{Matched asymptotic expansions}

The idea of the matched asymptotic expansions method is that one can find perturbative solutions to given problems in a black hole space-time by working in a regime in which one can separately solve the equations in a near-horizon region close to the black hole, and in the asymptotic region far away from the black hole.  This requires working in a regime in which there is an overlap region where both the near-horizon and asymptotic approximations are valid such that one can match the solutions there \cite{Poisson:2003nc}. 

The problem at hand, namely solving the equations for force-free electrodynamics \eqref{FFE_eqs} in a small $\alpha$ regime, perfectly fits into the framework of matched asymptotic expansions. 

Firstly, we have the region  \eqref{NH_region} where the perturbations in $\alpha$ makes sense. This is the {\sl near-horizon region} in our problem. In this region one should impose the condition of regularity of $\psi$ and $B^\phi$ at the event horizon $r=r_+$. 

Secondly, we have the {\sl asymptotic region} defined by
\begin{equation}
\label{asymptotic_region}
r \gg r_0
\end{equation}
in which we are far away from the event horizon of the Kerr black hole. In this region one should impose finiteness of $\psi$, $\Omega$ and $I$ for $r/r_0\rightarrow \infty$. Hence, we expand the fields $\psi$, $\Omega$ and $I$ as
\begin{equation}
\label{asymptotic_exp}
\begin{array}{c}\ds
\psi(r,\theta) = \psi_{(\cdot,0)} (\theta) + \frac{r_0}{r} \psi_{(\cdot,1)} (\theta) + \CO \Big(\frac{r_0^2}{r^2} \log \frac{r}{r_0}\Big)\,,
\\[2mm] \ds
r_0 \Omega(r,\theta) = \Omega_{(\cdot,0)} (\theta) + \frac{r_0}{r} \Omega_{(\cdot,1)} (\theta) + \CO \Big(\frac{r_0^2}{r^2}\log \frac{r}{r_0}\Big)\,,
\\[2mm] \ds
r_0 I(r,\theta) = I_{(\cdot,0)} (\theta) + \frac{r_0}{r} I_{(\cdot,1)} (\theta) + \CO \Big(\frac{r_0^2}{r^2}\log \frac{r}{r_0}\Big)\,,
\end{array}
\end{equation}
as well as
\begin{equation}
\label{asymp_B}
B^\phi(r,\theta) = \frac{B_{(\cdot,0)} (\theta) + \frac{r_0}{r} B_{(\cdot,1)} (\theta) + \CO \Big(\frac{r_0^2}{r^2}\log \frac{r}{r_0}\Big)}{r_0r^2}\,.
\end{equation}
One can have corrections that involves logarithms of $r_0/r$. However, such corrections do not show up in the analysis below to the order we are working. In the above expansion we are holding $\alpha$ fixed.

Thirdly, and finally, we have the {\sl overlap region}. This is defined by the intersection of the near-horizon and asymptotic regions
\begin{equation}
\label{overlap_region}
r_0 \ll r \ll \frac{r_0}{\alpha}\,.
\end{equation}
We notice that this region is well-defined thanks to being in the regime of small $\alpha$. 
In this region one can both expand in $\alpha$ and $r_0/r$ which means we have the possibility of connecting the solutions found in the near-horizon and asymptotic regions. Thus, in particular for $\psi(r,\theta)$ we have a double expansion in both $\alpha$ and $r_0/r$
\begin{equation}
\label{double_A}
\begin{aligned}
\psi (r,\theta) &= \psi_{(0,0)}(\theta) + \alpha^2 \psi_{(2,0)} (\theta) + \frac{r_0}{r} \psi_{(0,1)} (\theta) 
\\ & \ \ \ + \alpha^2 \frac{r_0}{r} \psi_{(2,1)}(\theta) + \CO (\alpha^4) + \CO\Big( \frac{r_0^2}{r^2} \log \frac{r}{r_0}\Big)\,,
\end{aligned}
\end{equation}
and similarly for $\Omega(r,\theta)$, $I(r,\theta)$ and $B^\phi(r,\theta)$. In detail, $\psi_{(m,n)}(r,\theta)$ refers to the function  multiplying $\alpha^m (r_0/r)^n$. The $r$ dependence of $\psi_{(m,n)}(r,\theta)$ is because it can possibly include a finite series in $\log (r/r_0)$. Thus, we require $\psi_{(m,n)}/r\rightarrow 0$ for $r\rightarrow \infty$. For $n=1$ we assume that there are no logarithmic terms since this is consistent with the small $\alpha$ analysis above.

The idea of the matched asymptotic expansions is then to first solve the stream equation in the near horizon region to order $\alpha$. The solution is also valid in the overlap region, therefore we use it as an input for the asymptotic region, where we solve the stream equation to leading order in the $r_0/r$ expansion. In this way we find a solution to order $\alpha$ that is valid up to infinity. Then one repeats this procedure to order $\alpha^2$ and so on.

\subsection{Asymptotic expansion at zeroth and first order}

Before turning to the matched asymptotic expansions of the Blandford-Znajek monopole we first examine the leading order part of the asymptotic region \eqref{asymptotic_region} using the expansions \eqref{asymptotic_exp}.
At zeroth order in $r_0/r$ the Stream equation \eqref{stream} simplifies to
\begin{equation}
\label{starteq1}
\sin \theta \Omega_{(\cdot,0)} \frac{d}{d\theta} \left( \sin \theta \Omega_{(\cdot,0)} \frac{d\psi_{(\cdot,0)}}{d\theta} \right) = I_{(\cdot,0)} \frac{dI_{(\cdot,0)}}{d\psi_{(\cdot,0)}}\,,
\end{equation}
where only the leading terms in \eqref{asymptotic_exp} contribute.
It is convenient to define a new variable $z(\theta)$ by
\begin{equation}
dz = \frac{d\theta}{\sin \theta \Omega_{(\cdot,0)}(\theta)}\,.
\end{equation}
Then \eqref{starteq1} becomes
\begin{equation}
\Big(\frac{d\psi_{(\cdot,0)}}{dz}\Big)^2  =   I_{(\cdot,0)}^2 + \mbox{const.}
\end{equation}
We impose that $I=0$ for $\theta=0$ as a boundary condition (can be derived from regularity of $B^\phi$ at $\theta=0$). Thus, the constant is required to be zero, and we deduce
\begin{equation}
\frac{d\psi_{(\cdot,0)}}{dz} = s I_{(\cdot,0)} \spa s = \pm 1\,,
\end{equation}
where we introduced the sign $s$.
This is equivalent to
\begin{equation}
\label{A0_eq}
\sin \theta \Omega_{(\cdot,0)} \frac{d\psi_{(\cdot,0)}}{d\theta}  =  s I_{(\cdot,0)}\,.
\end{equation}
Thus, given $\Omega_{(\cdot,0)}(\theta)$ and $\psi_{(\cdot,0)}(\theta)$ one can find $I_{(\cdot,0)}(\theta)$. 
One finds furthermore
\begin{equation}
\label{B_asy}
B_{(\cdot,0)} = - s \Omega_{(\cdot,0)}\,.
\end{equation}

Consider now the first order terms in $r_0/r$ in the expansions \eqref{asymptotic_exp} and \eqref{asymp_B}.  The integrability conditions \eqref{intcond1} and \eqref{intcond2} are
\begin{equation}
\label{A1_int}
\begin{array}{c}\ds
\Omega_{(\cdot,1)} \frac{d\psi_{(\cdot,0)}}{d\theta} = \psi_{(\cdot,1)} \frac{d\Omega_{(\cdot,0)}}{d\theta} \,,
\\[3mm]\ds
I_{(\cdot,1)} \frac{d\psi_{(\cdot,0)}}{d\theta} = \psi_{(\cdot,1)} \frac{dI_{(\cdot,0)}}{d\theta} \,,
\end{array}
\end{equation}
which gives $\Omega_{(\cdot,1)}$ and $I_{(\cdot,1)}$ in terms of $\psi_{(\cdot,1)}$. Using this with the Stream equation \eqref{stream} one gets the following equation for $\psi_{(\cdot,1)}(\theta)$
\begin{equation}
\label{A1_eq}
\frac{2\psi_{(\cdot,1)}}{\sin^2\theta}  \frac{d}{d\theta} \left(\sin^2 \theta\cos\theta  \Omega_{(\cdot,0)}^2 \right)= \frac{d}{d\theta} \left(\sin \theta  \Omega_{(\cdot,0)}^2 \frac{d\psi_{(\cdot,1)}}{d\theta} \right)  .
\end{equation}
Finally, one finds the following general expression for $B_{(\cdot,1)}$
\begin{equation}
\label{B1_eq}
B_{(\cdot,1)} = - (1+s) \Omega_{(\cdot,0)}   - s \psi_{(\cdot,1)} \frac{ \frac{d}{d\theta}\left( \sin^2 \theta \Omega_{(\cdot,0)}\right)}{\sin^3\theta} \,.
\end{equation}

\section{Failure of matched asymptotic expansions}

Our starting point in the near-horizon region \eqref{NH_region} is the expansions \eqref{alpha_exp} of $\psi$, $\Omega$ and $I$ in powers of $\alpha$. At order $\alpha^0$ we have the static monopole solution in the background of a Schwarzschild black hole \eqref{zero}. At first order in $\alpha$ we have the condition \eqref{alpha_one} that fixes $B^\phi$ at this order to \eqref{B_first}. Going to the overlap region \eqref{overlap_region} with expansions of the type \eqref{double_A} this fixes
\begin{equation}
\begin{array}{c} \ds
\psi_{(0,0)} = 1-\cos \theta \spa \psi_{(0,1)}=0 \,,
\\[3mm] \ds
I_{(1,0)}=\frac{1-2\Omega_{(1,0)}}{2} \sin^2\theta \spa I_{(1,1)}=\Omega_{(1,1)}=0 \,,
\\[3mm] \ds
B_{(1,0)} = - \frac{1-2\Omega_{(1,0)}}{2} \spa 
B_{(1,1)} = - \frac{1}{2} \,,
\end{array}
\end{equation}
For the asymptotic region \eqref{asymptotic_region} this means that at leading order in the $r_0/r$ expansion we have
\begin{equation}
\label{AOIB_0}
\begin{array}{c} \ds
\psi_{(\cdot,0)} = 1-\cos \theta + \CO(\alpha^2) \,,
\\[3mm] \ds
I_{(\cdot,0)}= \left( \frac{\alpha}{2}-\Omega_{(\cdot,0)} \right) \sin^2\theta + \CO(\alpha^3) \,,
\\[3mm] \ds
B_{(\cdot,0)} = - \left(\frac{\alpha}{2}-\Omega_{(\cdot,0)} \right) + \CO(\alpha^3) \,,
\end{array}
\end{equation}
and at first order in $r_0/r$ we have
\begin{equation}
\label{AOIB_1}
\begin{array}{c} \ds
\psi_{(\cdot,1)} =  \CO(\alpha^2) \,,
\spa 
\Omega_{(\cdot,1)}=  \CO(\alpha^3) \\[3mm] \ds
I_{(\cdot,1)}=  \CO(\alpha^3) \,,
\spa 
B_{(\cdot,1)} = - \frac{\alpha}{2}+ \CO(\alpha^3) \,.
\end{array}
\end{equation}
Considering the leading asymptotic part \eqref{AOIB_0} we see that Eq.~\eqref{A0_eq} is satisfied provided
\begin{equation}
\Omega_{(\cdot,0)}  (1+s) = s \frac{\alpha}{2} + \CO (\alpha^3) \,.
\end{equation}
This requires
\begin{equation}
\label{omega_cdot0}
s = 1 \spa \Omega_{(\cdot,0)}  =  \frac{\alpha}{4} + \CO (\alpha^3) \,.
\end{equation}
From this we get
\begin{equation}
\label{B_cdot0}
B_{(\cdot,0)} =  - \frac{\alpha}{4} + \CO(\alpha^3) \,,
\end{equation}
which is consistent with Eq.~\eqref{B_asy}. Note also that \eqref{A1_int}, \eqref{A1_eq} and \eqref{B1_eq} are consistent with \eqref{AOIB_1} provided $\Omega_{(\cdot,1)}= \CO(\alpha^5)$.

In the overlap region we get from \eqref{omega_cdot0} and \eqref{B_cdot0} that
\begin{equation}
\label{overlap_order_alpha}
\Omega_{(1,0)} = \frac{1}{4} \spa I_{(1,0)}=\frac{1}{4} \sin^2\theta
\spa
B_{(1,0)} = - \frac{1}{4} \,.
\end{equation}
In the near-horizon region \eqref{NH_region} this reproduces what we already found in \eqref{Omega_1cdot} at second order in $\alpha$. However, notice that we did this without invoking the second order in $\alpha$. Thus, even if we found the same result, the method is completely different. 

We conclude from the above that up to first order in $\alpha$ and first order in $r_0/r$, the Blandford-Znajek monopole solution is consistent with the matched asymptotic expansions analysis.

\subsection{Inconsistency at order $\alpha^2 r_0/r$}

There are no further terms at zeroth and first order in $\alpha$ than what we already discussed. Thus, the first new term that we encounter is at order $\alpha^2 r_0/r$. We now study first what we can infer from the near-horizon region \eqref{NH_region} about a term of this order, and subsequently what we can infer from the asymptotic region \eqref{asymptotic_region}.

For the near-horizon region we can use the analysis already reviewed above. This corresponds to the analysis of $\alpha^2$ corrections of the original paper of Blandford and Znajek \cite{Blandford:1977ds} which is reproduced in subsequent papers \cite{McKinney:2004ka,Tanabe:2008wm,Pan:2015haa}. Since we concluded that \eqref{Omega_1cdot} holds, we can use this in \eqref{S2gen} to get \eqref{fixedS2}.
The solution to this obeying that $\psi$ and $B^\phi$ are regular at the event horizon is Eqs.~\eqref{a2}-\eqref{r} that gives $\psi_{(2,\cdot)}$. At large $r/r_0$ this gives \eqref{A2_larger}. From this we get the following prediction in the overlap region for the term in $\psi$ at order $\alpha^2 r_0/r$
\begin{equation}
\label{NH_A21}
\psi_{(2,1)} = \frac{1}{8} \sin^2 \theta \cos\theta \,.
\end{equation}

For the asymptotic region we can use Eq.~\eqref{A1_eq} for $\psi_{(\cdot,1)}$. Write
\begin{equation}
\psi_{(\cdot,1)} = \alpha^2 \psi_{(2,1)} + \CO(\alpha^4) \,.
\end{equation}
Then Eq.~\eqref{A1_eq} gives
\begin{equation}
\label{A21_eq}
\left[ \frac{d^2}{d\theta^2} + \cot \theta \frac{d}{d\theta} -4\cot^2\theta+2\right] \psi_{(2,1)} = 0 \,.
\end{equation}
One can easily check that the prediction \eqref{NH_A21} does not obey this equation. In fact, 
the general solution to Eq.~\eqref{A21_eq} is a linear combination of $\sin^2 \theta$ and a function of $\theta$ that diverges at $\theta\rightarrow 0$ (see also \footnote{The same behaviour at large $r$ has been found also in \cite{Li:2017qzu}.}). Therefore, the near-horizon region and the asymptotic region are not consistent with each other in the overlap region \eqref{overlap_region}. Thus, we have found that the Blandford-Znajek monopole leads to inconsistencies at order $\alpha^2 r_0/r$.

In conclusion, we have shown that the Blandford-Znajek monopole does not exist for small $\alpha$ since assuming its existence leads to inconsistencies.


\section{Conclusion and Outlook}

In this Letter, we have considered the Blandford-Znajek (split-)monopole solution originally defined in \cite{Blandford:1977ds} in terms of a perturbative expansion in $\alpha$, the rotation parameter of the Kerr black hole.
We have formulated the criterion that since one defines the Blandford-Znajek monopole as a perturbation in $\alpha$  of a monopole solution around the Schwarzschild black hole, then the perturbations should be small everywhere outside the event horizon. Imposing this criterion we have found that the perturbative construction of the Blandford-Znajek monopole is inconsistent. This is revealed already at order $\alpha^2$.

The results of this Letter are supported by our forthcoming paper \cite{New} where we perform a general analysis of the FFE equations in the background of a Kerr black hole. The main idea of \cite{New} is to consider these equations close to the rotation axis and demand that solutions are regular at the rotation axis. This is seen to provide an alternative argument for the result of this Letter.

As mentioned in the introduction, there are several papers studying the Blandford-Znajek monopole numerically \cite{Komissarov:2001sjq,Komissarov:2004ms,McKinney:2004ka,McKinney:2006sc,Tchekhovskoy:2009ba,Palenzuela:2010xn,Contopoulos:2012py,Nathanail:2014aua,Mahlmann:2018ukr}. It would be highly interesting to consider how the result of this Letter can be in accordance with these studies. One way to pursue this would be to have a more close comparison between the numerical solutions for small $\alpha$ and the analytical results of this Letter and our forthcoming paper \cite{New}. One option could be that the numerical solutions do not obey the correct boundary conditions, corresponding to the ones outlined in this Letter. Another option, that we find more likely, is that the numerical solutions correspond to a physically different branch of solutions, thus not connected to the static monopole solution at $\alpha=0$. Potentially, this means that if we start with a different solution to the FFE equations in the background of a Schwarzschild black hole, but a solution that still asymptotes to the static monopole solution far away from the black hole, one could find a well-defined solution for small $\alpha$ using matched asymptotic expansions \cite{New}. Presumably, it would then follow that this is the solution that has been found numerically, and not the original Blandford-Znajek monopole. To verify this statement one needs further work, both on the numerical and analytical side.

We note that a comparison between analytical and numerical solutions at small $\alpha$ is potentially challenging. The numerical studies need to impose regularity across the light-sheet surfaces, and if $\alpha$ is small then the first light-sheet surface is very close to the event horizon of the Kerr black hole. This means that one needs a high resolution in the numerics to obtain a sufficiently accurate description of the electromagnetic configuration between these two surfaces \cite{Contopoulos:2012py,Nathanail:2014aua}. 

In light of the findings of this Letter, as well as our forthcoming publication \cite{New}, it would be interesting to consider also other perturbative constructions of solutions of the FFE equations in the background of a slowly rotating Kerr black holes. This includes the parabolic, vertical uniform and hyperbolic solutions in the Schwarzschild background \cite{Blandford:1977ds,Pan:2014bja,Gralla:2015vta} as well as the Blandford-Znajek monopole for Schwarzschild with a cosmological constant \cite{Wang:2014vza}.

\begin{acknowledgments}
We thank Maria Rodriguez for useful comments and for reading the manuscript of this Letter. We thank Jay Armas, Roberto Bruschini, Alfredo Glioti and Niels Obers for useful discussions. T.~H.~acknowledges support from the Independent Research Fund Denmark grant number DFF-6108-00340 ``Towards a deeper understanding of black holes with non-relativistic holography". M. O.~acknowledges support from the project ``HOLOgraphic description of strongly coupled SYStems (HOLOSYS)" financed by Fondo Ricerca di Base 2015 of the University of Perugia. T.~H.~thanks Perugia University  and M.~O.~thanks Niels Bohr Institute for hospitality.
\end{acknowledgments}


%

\end{document}